\documentclass[runningheads]{llncs}

\usepackage[T1]{fontenc}
\usepackage{algorithm}%
\usepackage{algorithmicx}%
\usepackage{algpseudocode}%
\usepackage{amsmath,amssymb,amsfonts}%
\usepackage{booktabs}    
\usepackage{comment}
\usepackage{graphicx}
\usepackage{listings}%
\usepackage{longtable}
\usepackage{manyfoot}%
\usepackage{makecell}     
\usepackage{multirow}%
\usepackage{subcaption} 
\usepackage{textcomp}%
\usepackage[dvipsnames]{xcolor}
\usepackage{microtype}
\usepackage[textwidth=1.5in]{todonotes}

\definecolor{RevisionColor}{RGB}{88, 24, 124}

\usepackage[hidelinks]{hyperref}
\usepackage{color}

\lstdefinelanguage{XML}{
  basicstyle=\ttfamily\small,
  showstringspaces=false,
  breaklines=true,
  morestring=[b]",
  moredelim=[s][\bfseries\color{black}]{<}{>},
  morecomment=[s]{<!--}{-->},
}

\lstdefinelanguage{json}{
  basicstyle=\ttfamily\small,
  showstringspaces=false,
  breaklines=true,
  morestring=[b]",
  stringstyle=\color{black},
  morecomment=[l]{//},
}

\definecolor{SimpleC}{HTML}{2E7D32}   
\definecolor{MediumC}{HTML}{EF6C00}   
\definecolor{ComplexC}{HTML}{C62828}  

\begin{document}

\title{A Hybrid LLM-Based Framework for Automated Security Annotation Generation in Business Process Models}

\titlerunning{A Hybrid LLM-Based Framework for Automated Security Annotation}

\author{Md Kamrul Islam\inst{1}\orcidID{0009-0000-8052-491X}
\and Tiphaine Henry\inst{2}\orcidID{0000-0002-7981-8934} 
\and Mattia Salnitri \inst{3}\orcidID{0000-0002-9736-2774}
\and Julius Köpke \inst{4}\orcidID{0000-0002-6678-5731}
\and Sami Souihi \inst{2}\orcidID{0000-0002-0986-170X}
}

\authorrunning{Md Kamrul Islam et al.}

\institute{CentraleSupélec, Gif-sur-Yvette, France \email{mdkamrul.islam@student-cs.fr}
\and Université Paris-Saclay, Palaiseau, France \email{\{tiphaine.henry,sami.souihi\}@cea.fr}\\
\and University of Bergamo, Bergamo, Italy, \email{mattia.salnitri@unibg.it}\\  
\and University of Klagenfurt, Klagenfurt, Austria\\ \email{julius.koepke@aau.at}}
\maketitle              
\begin{abstract}
The modelling and analysis of secure business processes require the incorporation of security annotations into process models. Although BPMN extensions, including SecBPMN2, exist for this purpose, the derivation of accurate and complete security annotations from natural-language specifications remains a manual, expert-intensive, and error-prone task. This paper presents a hybrid framework that takes a BPMN process model and a security requirements document as input and automatically generates security annotations adhering to the SecBPMN2 specification. The approach combines Large Language Model (LLM)--based semantic extraction with schema-constrained mapping, rule-based normalization, and deterministic validation.
The framework is evaluated comprehensively on a curated dataset of 27 process models from various domains. The results indicate that it consistently produces structurally valid SecBPMN2 annotations with high schema completeness. Compared to human security analysts, the system achieves substantially higher precision (0.58 vs.\ 0.29) while maintaining comparable recall (0.52 vs.\ 0.50) and reduces erroneous or misplaced annotations by nearly 50\%. In addition, annotation generation is significantly faster than manual annotation.
These findings demonstrate that hybrid LLM- and rule-based automation can reduce modeling effort while improving consistency and reliability, thereby providing a scalable foundation for security-by-design BPM.

\end{abstract}

\section{Introduction}
BPMN 2.0\footnote{\url{https://www.omg.org/spec/BPMN}} is the established industry standard notation for modeling business processes across the BPM lifecycle. However, BPMN lacks native support for expressing security requirements, which is increasingly problematic in regulated and data-sensitive
environments~\cite{chinosi2012bpmn}. To address this limitation, several extensions-including SecureBPMN~\cite{brucker2012securebpmn}, SecBPMN~\cite{salnitri2017designing}, and PE-BPMN~\cite{pullonen2019privacy} have been proposed to enable security-by-design by introducing explicit security annotations at the process-modeling level. Despite their expressive power, these extensions have seen limited adoption in practice. This limited adoption is largely due to the fact that security annotations are typically derived manually from natural-language descriptions, a task that requires specialized expertise, is prone to inconsistency, and scales poorly in complex and evolving regulatory contexts~\cite{ramadan2020semi}.

Recent advances in LLMs have demonstrated strong capabilities in interpreting unstructured text and generating structured process representations~\cite{kopke2024efficient,nour2024nala2bpmn}. However, security-annotated process modeling poses additional challenges, as security requirements are often implicit, context-dependent, and linguistically heterogeneous. In this work, we focus on SecBPMN2, which extends BPMN with semantically grounded vocabulary of security annotations and precisely defined attachment rules, enabling explicit and verifiable specification of properties such as confidentiality, integrity, accountability, and separation of duties~\cite{salnitri2017designing}. 

While these constraints are essential for systematic validation, their rich semantics imposes extensive effort for human modelers and substantial challenges for LLM-based text-to-model generation. These challenges motivate the following research question.
\begin{quote}
\textbf{RQ.} \emph{How can an LLM augmented system automatically annotate BPMN models with SecBPMN2-compliant security annotations based on textual process descriptions?}
\end{quote}

To answer this question, this paper makes three contributions. First, we propose a hybrid automation pipeline that derives SecBPMN2 annotations from natural-language descriptions by coupling LLM-based extraction and mapping with rule-based normalization and validation. Second, we introduce a benchmark dataset aligning textual process descriptions with expert-designed SecBPMN2 ground-truth models. Finally, we present an empirical evaluation across multiple LLMs and prompting strategies, including Retrieval-Augmented Generation (RAG), using quantitative metrics and human baselines. Together, these contributions advance LLM-based support for security-aware process modeling by reducing manual effort, improving consistency, and enabling the systematic operationalization of SecBPMN2 annotations.

The remainder of the paper is organized as follows. Section~\ref{relworks} reviews related work, Section~\ref{motex} presents a motivating example, and Section~\ref{approach} describes the proposed automation pipeline. Section~\ref{sec:implementation} outlines the prototype implementation, Section~\ref{experiments} reports the experimental evaluation, and Section~\ref{conclusion} concludes the paper.

\section{Related Work}
\label{relworks}

\begin{table}[t]
\caption{Comparison of related work (NL = natural language).}
\label{tab:relatedwork-comparison}
\centering
\begin{tabular}{p{3.5cm} p{1cm} p{1.5cm} p{1.9cm} p{1.5cm} p{2.0cm}}
\hline
\textbf{Approach} &
\textbf{NL Input} &
\textbf{Auto-mation} &
\textbf{Security Semantics} &
\textbf{Process Integration} &
\textbf{Constraint Enforcement} \\
\hline

Security extensions \cite{brucker2012securebpmn,pullonen2019privacy,cherdantseva2013reference,salnitri2017designing} 
& No & Manual & Explicit & BPMN / SecBPMN2 & Yes \\

Rule-based generation \cite{friedrich2011process} 
& Yes & Partial & None & BPMN & No \\

LLM-based generation \cite{kourani2024process,kourani2025evaluating,kopke2024efficient,nour2024nala2bpmn} 
& Yes & Yes & None & BPMN & No \\

Security req. extraction \cite{van2024nlp4pbm,siavvas2024digital} 
& Yes & Yes & Explicit (text) & None & Partial \\

\textbf{This work} 
& Yes & Yes & Explicit (SecBPMN2) & BPMN / SecBPMN2 & Yes \\

\hline
\end{tabular}
\end{table}

Research on security-aware BPM spans three main directions: BPMN security extensions, automated BPMN generation from text, and extraction of structured security requirements. These directions differ along five dimensions- n natural-language input, automation, security semantics, integration with BPMN artifacts, and constraint enforcement-as summarized in Table~\ref{tab:relatedwork-comparison}.

The first line of research extends BPMN with explicit security constructs. BPMN primarily captures functional behavior and provides limited support for security and privacy concerns~\cite{chinosi2012bpmn}. Extensions address this gap: SecureBPMN introduces authorization and duty constraints~\cite{brucker2012securebpmn}, and PE-BPMN focuses on privacy and data protection~\cite{pullonen2019privacy}. SecBPMN and SecBPMN2 provide a comprehensive framework for security modeling. SecBPMN2-ml defines a structured vocabulary of security goals grounded in RMIAS~\cite{cherdantseva2013reference,salnitri2017designing}, while SecBPMN-Q enables formal policy verification. These approaches provide explicit semantics, tight BPMN integration, and constraint enforcement, but rely on manual annotation and do not support natural-language input.

The second line focuses on automating BPMN generation from textual descriptions. Early approaches rely on syntactic parsing and rule-based extraction~\cite{friedrich2011process}, while recent work leverages large language models, including translation pipelines~\cite{kourani2024process}, iterative prompting~\cite{kourani2025evaluating}, conversational modeling~\cite{kopke2024efficient}, and direct BPMN XML generation~\cite{nour2024nala2bpmn}. These methods support natural-language input and automated model construction, but focus on functional structure and do not incorporate explicit security semantics or constraint enforcement.

The third line concentrates on extracting structured security requirements from natural-language specifications. Empirical studies show that LLMs can identify entities and relations but struggle to construct complete and semantically consistent formal models under controlled benchmarks~\cite{van2024nlp4pbm}. Domain-specific approaches derive structured requirements from standards such as ISO/IEC~27001~\cite{siavvas2024digital}. While these methods produce explicit security representations at the textual level, they operate largely independently of BPMN artifacts and provide limited process-level constraint enforcement.

In summary, existing work addresses either security semantics, automation from text, or structured requirement extraction. No existing approach integrates natural-language input, automated generation, SecBPMN2 semantics, BPMN-level integration, and systematic constraint enforcement within a framework.

\section{Motivating Example}
\label{motex}

\begin{figure}[t]
\centering

\newsavebox{\rbtbox}
\sbox{\rbtbox}{%
\includegraphics[width=0.78\textwidth]{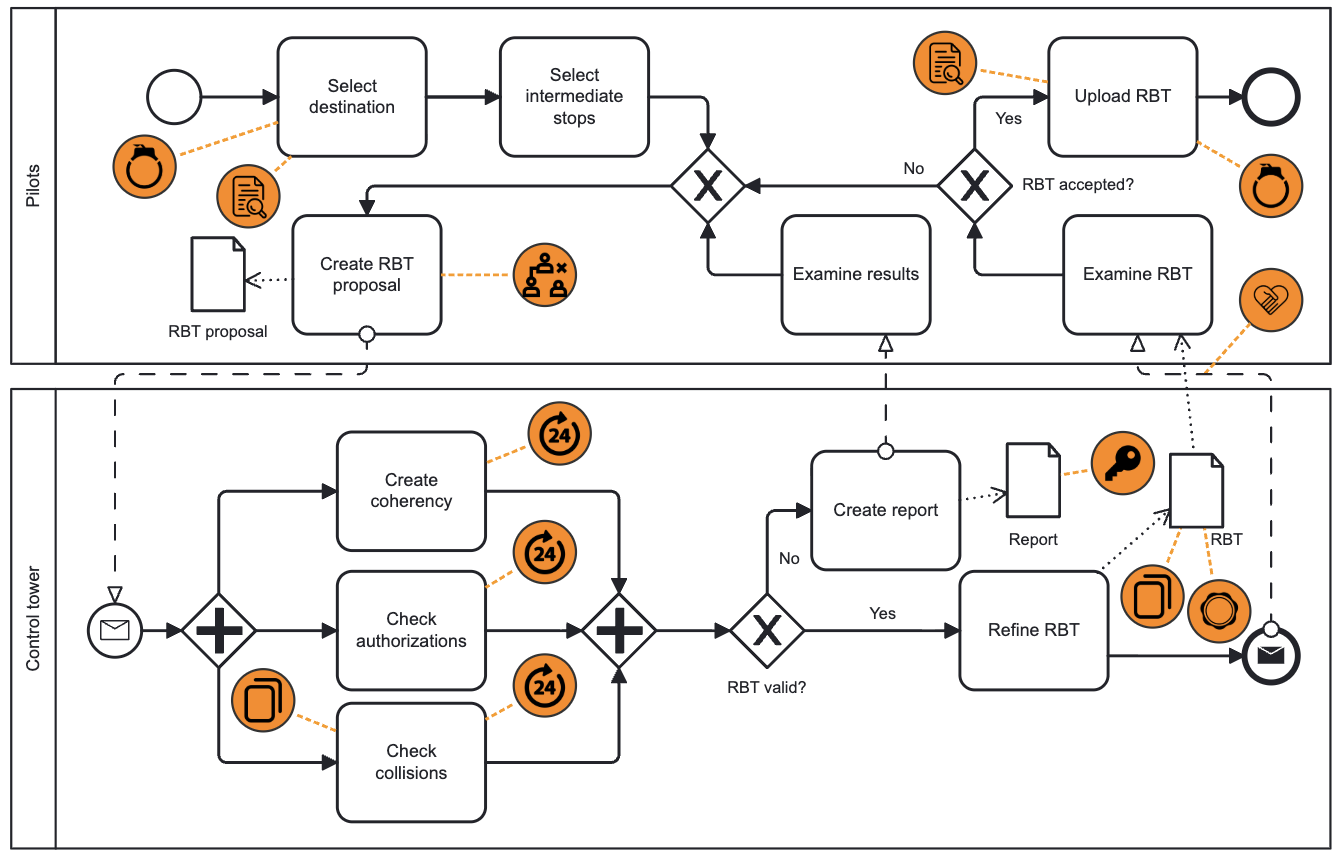}
}

\begin{minipage}[t]{0.78\textwidth}
    \centering
    \usebox{\rbtbox}
\end{minipage}
\hfill
\begin{minipage}[t]{0.20\textwidth}
    \centering
    \includegraphics[height=\ht\rbtbox]{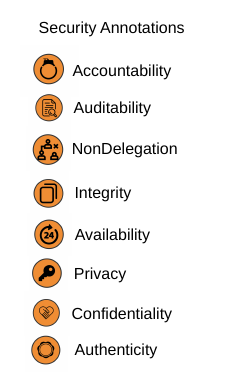}
\end{minipage}

\caption{Simplified RBT negotiation process annotated with SecBPMN2~\cite{salnitri2017designing}.}
\label{fig:rbt-model}

\end{figure}

Automated security annotation is particularly important in settings that involve sensitive or safety-critical information. A representative example is the Route-Based Trajectory (RBT) negotiation process introduced by Salnitri et al.~\cite{salnitri2017designing}, which models the iterative exchange between pilots and air-traffic controllers during the proposal, revision, and approval of flight trajectories. Figure~\ref{fig:rbt-model} depicts a simplified version of this process and serves as the motivating example used throughout this paper.
The RBT process exchanges highly sensitive trajectory data and therefore requires confidentiality of flight messages, authenticity of proposed and approved trajectories, accountability of approving actors, auditability of critical decisions, and continuous availability of safety checks. In practice, these requirements are expressed informally (e.g., restricting message access, logging approvals, or ensuring continuous collision checks) and are not directly translatable into formal security annotations. Mapping such requirements to SecBPMN2 requires identifying relevant BPMN elements, selecting appropriate security goals, and instantiating schema-valid parameters in accordance with SecBPMN2 attachment rules. Even for compact models, this manual translation requires substantial expertise and effort and is prone to inconsistency and omissions. 

The RBT example highlights a fundamental challenge in security-aware BPM. Security requirements are often expressed in flexible, implicit, and context-dependent natural language, whereas SecBPMN2 requires explicit, schema-constrained annotations attached to specific BPMN elements. 

\section{Approach}
\label{approach}

Building on the challenge identified in the motivating example, this section details the proposed approach for automatically generating SecBPMN2 security annotations from natural-language process descriptions. The approach takes as input (i) a syntactically valid BPMN~2.0 model and (ii) a natural-language description of security-relevant requirements. It produces a SecBPMN2-compliant process model in which security annotations are instantiated, validated, and attached to the appropriate BPMN elements. To bridge this gap, the approach decouples semantic interpretation from structural validation. LLM components perform security-goal extraction from text, while rule-based components enforce SecBPMN2 attachment rules, schema constraints, and structural consistency.

In the following, (i) \textit{security requirements} denote textual security requirements expressed in the input documentation; (ii) \textit{Security goals} denote the abstract security properties (e.g., confidentiality, integrity, or availability) derived from these requirements; (iii) \textit{Security annotations} denote the SecBPMN2 constructs that instantiate these goals and are attached to BPMN elements.

\subsection{Overview of the Hybrid Approach}
\label{subsec:approach-overview}

\begin{figure}[t]
\centering
\includegraphics[width=\linewidth]{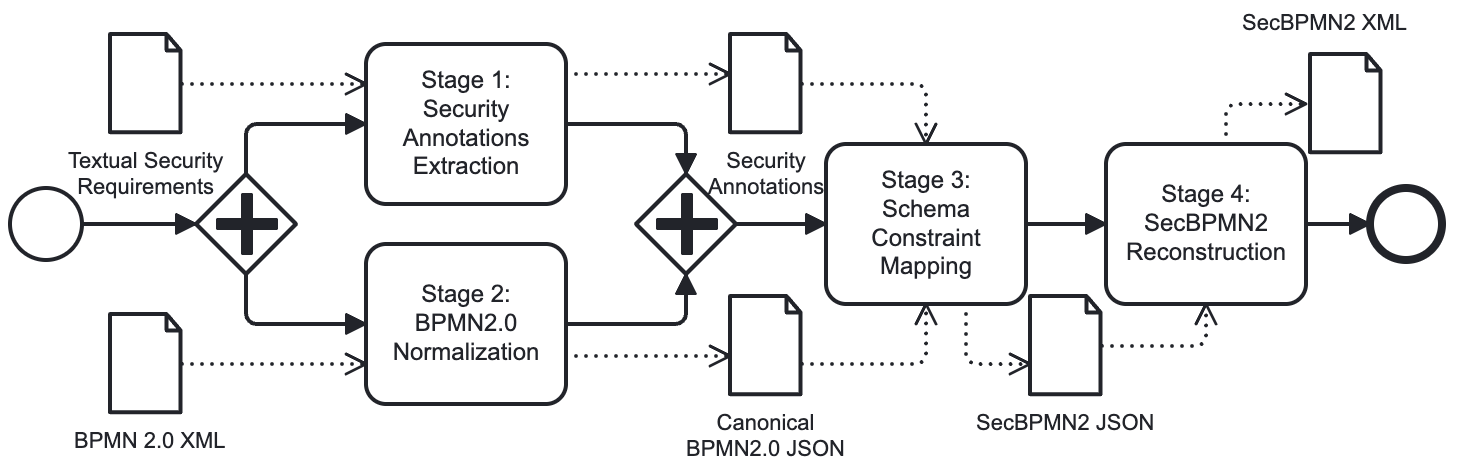}
\caption{Four-stage hybrid pipeline for deriving SecBPMN2 annotations from textual security requirements.
}
\label{fig:pipeline_overview}
\end{figure}

The proposed methodology is a four-stage pipeline, illustrated in Figure~\ref{fig:pipeline_overview}. Stages~1 and~2 operate in parallel. Stage~1 analyzes the natural-language process description to extract security goals without committing to concrete BPMN elements. Stage~2 transforms the BPMN~2.0 XML model into a normalized, graph-oriented JSON representation that exposes the structural primitives required for downstream reasoning. In Stage~3, the outputs of these two stages are combined through an LLM-assisted, schema-constrained mapping procedure that enforces SecBPMN2 attachment rules. Stage~4 then reconstructs the enriched model as SecBPMN2-compliant XML using rule-based templates.

The pipeline's design is guided by two principles. The semantic interpretation and structural validation are decoupled to prevent linguistic ambiguity from propagating into violations of SecBPMN2 annotations. 
The intermediate representations are reversible. This ensures the traceability from textual requirements to process models. The usage of JSON allows current LLMs to produce outputs that practically fully comply with the schemas \cite{kopke2024efficient}. However, the particular serialization format is an implementation issue that can be adopted based on the capabilities of particular LLMs.

Before detailing the individual stages, we formalize the modeling assumptions and annotation schema that define the scope and constraints of the approach.

\subsection{Modeling Assumptions and Security Annotation Schema}
\label{subsec:modeling_assumptions}

\paragraph{\textbf{Assumptions.}}
The method operates on BPMN~2.0 within the subset supported by SecBPMN2. The input model is assumed to be syntactically valid and to expose the BPMN elements relevant for security annotation (tasks, events, gateways, message flows, data objects, and organizational elements). 

The accompanying textual description is expected to reference these elements either directly or indirectly through actors, activities, communications, or information artifacts. The approach does not require references to match BPMN element names or types verbatim; for example, actor-centric descriptions such as ``pilots are accountable for selecting a destination`` can be aligned with the corresponding BPMN activity through semantic mapping. Semantic alignment is established during Stage~3 using textual anchors and process context. Security requirements are assumed to be expressible using SecBPMN2 annotation types~\cite{kopke2023designing}. Descriptions that lack any identifiable process-level anchor may lead to ambiguous mappings and reduced annotation accuracy.

\paragraph{\textbf{Schema and notation.}}
Let $J_{\mathrm{bpmn}}$ denote the normalized BPMN-JSON representation of the input model (produced in Stage~2), $\mathcal{L}$ the set of candidate security annotations extracted from text (Stage~1), and $J_{\mathrm{sec}}$ the resulting SecBPMN-JSON model after mapping and validation (Stage~3). The annotation schema enforced by our approach follows SecBPMN2-ML and represents a security annotation as the tuple $\alpha=(e,g,p)$, where $e \in \mathcal{E}$ denotes a BPMN element, $g$ a security goal, and $p$ a goal-specific parameter set. BPMN elements are partitioned as
$\mathcal{E}=\mathcal{E}_{\mathrm{Act}}\cup\mathcal{E}_{\mathrm{DO}}\cup
\mathcal{E}_{\mathrm{MF}}\cup\mathcal{E}_{\mathrm{GW}}\cup
\mathcal{E}_{\mathrm{ORG}}$. Schema-compliant security associations satisfy a compatibility predicate $\mathit{compat}(g,\mathit{type}(e))$, which enforces type-level consistency between security goals and BPMN element categories (e.g., confidentiality applies only to data objects and message flows). These definitions lay the ground for the algorithmic enforcement of the schema during mapping stage.

Each SecBPMN2 annotation type encodes its admissible BPMN element category through a suffix 
$s \in \{\mathrm{Act},\mathrm{DO},\mathrm{MF},\mathrm{GW},\mathrm{ORG}\}$, 
which corresponds to the respective partition of $\mathcal{E}$. This suffix determines the permissible target category $\mathrm{type}(e)$ for a constraint attached to a BPMN element.

\subsection{Stage 1: Security Annotations Extraction}
\label{sec:stage1-extraction}

Stage~1 derives a set of candidate security annotations $\mathcal{L}$ from the natural-language process description provided by the security expert. The objective is to identify security goals \(g\) (e.g., accountability, auditability, integrity) together with an inferred target \(\mathrm{type}(e)\in\mathcal{E}\) (e.g., activity, data object, message flow, gateway, participant), while avoiding the concrete BPMN~2.0 element identifiers. At this stage, extracted annotations are semantic and model-agnostic: each annotation encodes a security goal \(g\), goal-specific parameter placeholders \(p\), and textual anchors (e.g., referenced actions, data items, or actors), but is not bound to a specific BPMN element \(e\). This separation is a deliberate design choice motivated by the distinction between semantic interpretation and structural validation. Security requirements are often expressed independently of concrete BPMN element identifiers, whereas SecBPMN2 annotations must satisfy strict attachment constraints. By first extracting model-agnostic security annotations and postponing element-level attachment, the approach decouples semantic reasoning from structural mapping, enabling deterministic validation and reducing the risk of schema-incompatible annotations.

A role-based structured prompt combined with chain-of-thought reasoning and few-shot examples extracts a set of candidate security annotations $\mathcal{L}$ under type and parameter constraints~\cite{sultan2024scot}. The prompt guides the LLM through a structured socio-technical analysis aligned with STS-ML~\cite{paja2015modelling}. It extracts three perspectives: (i) a \textit{social view} describing actors, roles, and their relationships, (ii) an \textit{information view} identifying relevant data objects and exchanges, and (iii) an \textit{authorization view} capturing access rights and delegation assumptions relevant for extracting security annotations. It incorporates a security annotation catalog that maps each SecBPMN2 annotation type-identified by a suffix 
$s \in \{\mathrm{Act},\mathrm{DO},\mathrm{MF},\mathrm{GW},\mathrm{ORG}\}$ -
to its admissible BPMN element category in $\mathcal{E}$ and required parameter set $p$.

The output of Stage~1 is a schema-conformant JSON representation of $\mathcal{L}$. Each label specifies a candidate security goal $g$ with goal-specific parameters $p$, together with an inferred target category $\mathrm{type}(e)\in\mathcal{E}$ and associated textual anchors. For the RBT negotiation process, Stage~1 yields annotations corresponding to goals such as accountability, auditability, and integrity, instantiated as SecBPMN2 annotation types (e.g., \texttt{accountabilityAct}, \texttt{auditabilityAct}, \texttt{integrityAct}).

Multiple extraction configurations are supported, including alternative prompting strategies and retrieval-augmented extraction. When enabled, concise definitions from SecBPMN2 documentation are retrieved and incorporated into the prompt to improve alignment with the intended security semantics.

\subsection{Stage 2: BPMN~2.0 Normalization}
Stage~2 constructs the normalized BPMN-JSON representation $J_{\mathrm{bpmn}}$ from the BPMN~2.0 XML input. The normalization yields a canonical, token-efficient representation that serves as the structural basis for downstream processing. 

It retains all BPMN constructs, including tasks and events, gateways, message flows, data objects and data object references, as well as participants (pools), lanes, message definitions, and data associations. Each element is represented using a stable identifier and explicit connectivity (e.g., incoming and outgoing relations), while XML-specific syntactic overhead, such as namespaces, deeply nested tags, and presentation-oriented BPMN-DI layout metadata, is omitted. The original BPMN XML, including BPMN-DI, is preserved for visualization and export in later stages.

Beyond flattening element encodings, normalization reconstructs an explicit control-flow structure through a linear-time traversal of the BPMN control-flow graph encoded in the model. Gateways are represented as structured branching nodes with explicit paths and retained conditions. The resulting $J_{\mathrm{bpmn}}$ provides a compact, machine-oriented representation that reduces token overhead while preserving the identifiers and relations required for deterministic reconstruction and unambiguous label-to-element mapping in Stage~3, aligning with evidence that lightweight, text-centric encodings outperform verbose formats such as XML for LLM-based structured reasoning~\cite{yang2026structeval}. Although alternative graph-based representations could also be employed, the chosen format offers a practical balance between structural fidelity, reversibility, and integration with subsequent validation and reconstruction steps.

\subsection{Stage 3: Mapping Security Annotations to BPMN Elements}
\label{sec:stage3-mapping}

Stage~3 maps the security annotations $\mathcal{L}$ from Stage~1 to BPMN2.0 elements in the normalized BPMN-JSON representation $J_{\mathrm{bpmn}}$ (Stage2). Given $(J_{\mathrm{bpmn}}, \mathcal{L})$, the objective is to construct a SecBPMN2-JSON model $J_{\mathrm{sec}}$ that preserves the BPMN structure while augmenting it with security annotations (represented as \texttt{secConstraints}) and corresponding \texttt{secAssociations}. Each candidate annotation is instantiated as a concrete SecBPMN2 annotation $\alpha=(e,g,p)$ by associating it with a BPMN element identifier that satisfies the schema constraints defined in Section~\ref{subsec:approach-overview}.

Mapping is performed using a single low-temperature LLM invocation ($T=0.1$) followed by deterministic validation. The prompt consists of (i) the normalized BPMN-JSON representation, (ii) a compact index containing BPMN element identifiers, labels, types, and participant information, and (iii) the candidate security annotations extracted in Stage~1. The LLM is not tasked with generating a BPMN model; instead, it preserves the BPMN representation unchanged and generates only SecBPMN2 security constraints and their associations (Listing~\ref{lst:stage3_prompt}).

\begin{lstlisting}[frame=single,
                   basicstyle=\footnotesize\ttfamily,
                   caption={Representative Stage~3 prompt.},
                   label={lst:stage3_prompt}]
Input: BPMN Base, Compact BPMN Index, Security Labels

Task: Copy the BPMN Base unchanged. Map security labels to
schema-compatible BPMN element identifiers using participant
and type constraints. Generate only securityConstraints and
securityAssociations.
\end{lstlisting}

Since Stage~1 provides only textual anchors (e.g., references to actors, activities, data items, or communications), the LLM resolves paraphrases and semantic matches under BPMN type constraints and participant scoping. The resulting mappings are subsequently validated by checking target existence, attachment cardinality, and type compatibility through the predicate $\mathit{compat}(g,\mathit{type}(e))$. Any invalid mappings are removed, yielding a validated SecBPMN2-JSON model $J_{\mathrm{sec}}$. 

\begin{algorithm}[t] 
\caption{Stage~3: Security annotation mapping and validation.} 
\label{alg:stage3-mapper} 
\begin{algorithmic}[1] 
\Require BPMN model $J_{\mathrm{bpmn}}$, annotations $\mathcal{L}$, LLM $\mathsf{M}$, validator $\mathsf{V}$ 
\Ensure SecBPMN-JSON model $J_{\mathrm{sec}}$ 
\State $P \gets \mathrm{RenderPrompt}(J_{\mathrm{bpmn}}, \mathcal{L})$ 
\State $r \gets \mathrm{CallLLM}(\mathsf{M}, P)$ 
\State $J_{\mathrm{sec}} \gets \mathrm{ParseJSON}(r)$ 
\If{$J_{\mathrm{sec}}$ invalid}     
\State $J_{\mathrm{sec}} \gets \mathrm{FallbackMapping}(J_{\mathrm{bpmn}}, \mathcal{L})$ \EndIf 
\State $J_{\mathrm{sec}} \gets \mathsf{V}.\mathrm{ValidateAndFilter}(J_{\mathrm{sec}}, J_{\mathrm{bpmn}})$ 
\State \Return $J_{\mathrm{sec}}$ 
\end{algorithmic} 
\end{algorithm}  

\paragraph{\textbf{Example.}} In the RBT negotiation process (Fig.~\ref{fig:rbt-model}), a requirement such as the revision communication employs advanced encryption methods'' is extracted as a confidentiality annotation with an intended message-flow target. During Stage 3, the LLM maps the textual anchor ``revision communication'' to the corresponding BPMN message flow, while validation ensures that the resulting confidentiality annotation is attached only to a schema-compatible message-flow element. This combination of semantic mapping and deterministic validation yields structurally valid SecBPMN2 annotations.

\subsection{Stage 4: SecBPMN2 XML Reconstruction}
\label{sec:stage4-reconstruction}

Stage~4 reconstructs the validated SecBPMN-JSON model $J_{\mathrm{sec}}$ as SecBPMN2-compliant BPMN~2.0 XML. This stage restores the BPMN~2.0 and SecBPMN2 namespaces, materializes SecBPMN2 extension elements for each entry in \texttt{secCons\-traints} and \texttt{secAssociations}, and serializes the collaboration and process structures into a well-formed XML document. Because $J_{\mathrm{sec}}$ preserves the original BPMN element (e.g., tasks, gateways, message flows, participants), the resulting XML retains the same BPMN~2.0 control-flow and interaction structure as the input model, augmented with security annotations.

\section{Prototype Implementation}
\label{sec:implementation}
\begin{figure}[t]
  \centering
  \includegraphics[width=\textwidth]{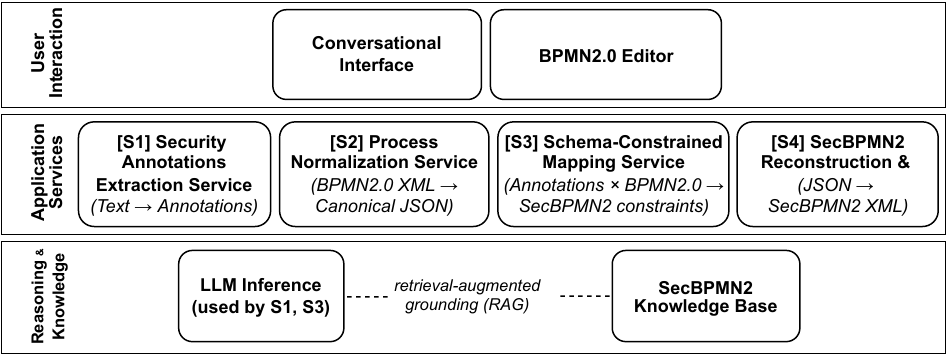}
  \caption{Architecture of the SecBPMN2 annotation prototype.}
  \label{fig:chatbot-architecture}
\end{figure}

To support reproducibility and demonstrate the practical realization of the proposed approach, we implemented a prototype that operationalizes the SecBPMN2 automation pipeline within a BPMN modeling environment. The prototype, extending an open-source BPMN assistant \footnote{\url{https://github.com/jtlicardo/bpmn-assistant}}, follows a three-layer architecture presented in Figure~\ref{fig:chatbot-architecture}. The user interaction layer comprises (1) a BPMN modeling interface that enables users to supply process models and (2) a conversational interface in which the user can provide the textual security requirements and inspect the SecBPMN2 annotations identified during Stage~1. To preserve compatibility with standard BPMN tools, the exported models retain the original BPMN structure and are augmented solely with SecBPMN2 extensions. The application services layer executes the four stages of the pipeline, including security annotation extraction, BPMN normalization, schema-constrained mapping, and SecBPMN2-compliant reconstruction. The reasoning and knowledge layer, which handles LLM inference requests, is invoked during security annotation extraction (Stage~1) and schema constraint mapping (Stage~3), while all structural validation and reconstruction steps are performed deterministically. The prototype implementation is publicly available to support reproducibility\footnote{Software artifact: \url{https://doi.org/10.5281/zenodo.19063772}}.

\section{Evaluation}
\label{experiments}

The evaluation assesses the effectiveness and practicality of the SecBPMN2 automation pipeline based on the generated SecBPMN2 artifacts. As the implementation is designed for experimentation and reproducibility, the evaluation focuses on artifact quality rather than interface usability. Specifically, it considers five objectives:
(E1) assessing whether the pipeline generates SecBPMN2 annotations that conform to the meta-model and attachment constraints across process complexity levels;
(E2) measuring how accurately the pipeline extracts and maps security annotations from natural-language specifications relative to expert-designed ground truth;
(E3) evaluating the quality and consistency of SecBPMN2 Assistant–generated annotations compared to security experts;
(E4) analyzing how extraction performance varies across different categories of security goals; and
(E5) assessing the computational cost of the pipeline and how runtime and token usage scale with process complexity.

\subsection{Experimental Setup}
\label{sec:exp-setup}

The evaluation is conducted on a curated dataset\footnote{Dataset: \url{https://doi.org/10.5281/zenodo.19064046}.} designed to align natural-language security requirements with SecBPMN2 process models. Because no public benchmark provides such paired artifacts, we constructed a dataset of 27 process models spanning aviation (9), healthcare (7), finance (6), public services (3), and hospitality (2). The models were collected from published SecBPMN2 case studies (20) and industrial examples (7). While several source models provided SecBPMN2 annotations, none included aligned natural-language security descriptions. We therefore manually authored one description per model by systematically translating BPMN constructs (e.g., participants, activities, gateways, message flows, and data objects) into domain-language narratives. The generated descriptions follow the assumptions introduced in \ref{subsec:modeling_assumptions} and were reviewed against the corresponding SecBPMN2 models to ensure coverage of all security-relevant process elements and annotated security requirements. Descriptions range from 97 to 670 words. Each dataset instance therefore consists of a BPMN process model, a natural-language security description, and a SecBPMN2 ground-truth annotation model, enabling reproducible evaluation of security annotation pipelines. Table~\ref{tab:secbpmn-dataset-summary-agg} summarizes the dataset characteristics by complexity tier (simple, medium, and complex), defined by the number of BPMN elements and security annotations.

We evaluate two extraction strategies: prompt-based and RAG, where prompts are enriched with passages retrieved from a knowledge base constructed from SecBPMN2, SecBPMN2BC, and STS-ML documentation~\cite{secbpmn2_reference,kopke2023designing,paja2015modelling}. 
Experiments are conducted with GPT-4.1-mini and Mistral Small 3.2 -- using identical decoding settings ($\text{temperature}=0.1$, $\text{max\_tokens}=30000$). The $\text{temperature}=0.1$ was used to balance reproducibility with robust annotation extraction and mapping by reducing stochastic variation while retaining limited generation flexibility. Retrieval follows a fixed hybrid BM25--FAISS configuration~\cite{johnson2019billion}, with $\text{top-}k=16$ and fusion weight $\alpha=0.35$ for BM25 and $1-\alpha$ for FAISS. Unless restricted to a single document type, we allocate retrieved context across sources (70\% SecBPMN2, 20\% SecBPMN2BC, 10\% STS-ML). FAISS embeddings are computed using \texttt{text-embedding-3-small}. All generated models are validated against the SecBPMN2 schema and compared with annotations produced by three participants: two experienced cybersecurity professionals and one Master’s student in cybersecurity, none of whom had prior SecBPMN2 experience.

The evaluation operationalizes the objectives defined in Section~\ref{experiments} using six metrics. Structural validity evaluates E1 by measuring conformance to the SecBPMN2 meta-model and correctness of annotation attachment. Precision, recall, and F1 score measure E2 and E3 by quantifying the correctness and completeness of extracted and mapped security annotations relative to expert annotations~\cite{kopke2024efficient,kourani2025evaluating}. For evaluation purposes, each security annotation is represented by three components: (i) the security goal, (ii) the SecBPMN2 annotation type, and (iii) the BPMN element to which the annotation is attached. Precision, recall, and F1 are computed using strict exact matching: a generated annotation is counted as a true positive only if all three components match a ground-truth annotation exactly. Generated annotations without a matching ground-truth annotation are counted as false positives, whereas ground-truth annotations that are not generated are counted as false negatives. Consequently, these metrics assess not only structural validity but also semantic correctness with respect to expert-designed SecBPMN2 annotations, as incorrectly inferred security goals, misplaced annotations, and hallucinated annotations are reflected as false positives or false negatives. Category-level recall supports E4 by analyzing extraction performance across different security goal classes. Finally, computational efficiency and reproducibility address E5 by measuring runtime cost, annotation effort, and stability of results under LLM nondeterminism~\cite{nour2024nala2bpmn}.

To address E1–E5, we conduct three experiments: (i) structural validity and extraction performance, (ii) human baseline and category-level analysis, and (iii) computational cost and scalability. All experiments use identical decoding parameters for comparability.

\begin{table}[t]
\caption{Dataset summary aggregated by complexity tier. Percentages report the distribution of security targets.}
\label{tab:secbpmn-dataset-summary-agg}
\centering
\small
\renewcommand{\arraystretch}{1.2}
\begin{tabular}{lccccccc}
\toprule
\textbf{Tier} & \textbf{\#} & \textbf{BPMN elems.} & \textbf{Sec. ann.} & \textbf{ACT} & \textbf{DO} & \textbf{MF} & \textbf{GW/ORG} \\
\midrule
Simple  
& 10 
& 9.7 (4--17) 
& 3.1 (1--6) 
& 87.1\% 
& 3.2\% 
& 9.7\% 
& 0.0\% \\

Medium  
& 7  
& 24.4 (11--38) 
& 4.0 (1--8) 
& 35.7\% 
& 46.4\% 
& 17.9\% 
& 0.0\% \\

Complex 
& 9  
& 54.0 (29--116) 
& 23.7 (7--64) 
& 46.9\% 
& 31.5\% 
& 18.8\% 
& 2.8\% \\
\bottomrule
\end{tabular}
\end{table}

\subsection{Results}
\label{sec:results}

This subsection reports the results with respect to the evaluation objectives defined in Section~\ref{experiments}. Key results are summarized in Table~\ref{tab:performance_and_human_baseline}.

\begin{table}
\caption{Structural validity, extraction performance, and human baseline.}
\label{tab:performance_and_human_baseline}

\centering
\small

\begin{subtable}[t]{\textwidth}
\centering
\caption{Schema validity after mapping, aggregated by model and complexity tier.}
\label{tab:schema_validity_by_complexity}
\begin{tabular}{lcc|cc}
\toprule
\multirow{2}{*}{\textbf{Complexity}} 
& \multicolumn{2}{c}{\textbf{Prompt-based}} 
& \multicolumn{2}{c}{\textbf{RAG}} \\
\cmidrule(lr){2-3} \cmidrule(lr){4-5}
& GPT & Mistral & GPT & Mistral \\
\midrule
Simple (avg.)  & 0.926 & 0.954 & \textbf{0.987} & 0.905 \\
Medium (avg.)  & 0.801 & 0.857 & \textbf{0.911} & 0.808 \\
Complex (avg.) & 0.900 & 0.897 & \textbf{0.924} & 0.865 \\
\bottomrule
\end{tabular}
\end{subtable}
\vspace{0.75em}

\begin{subtable}[t]{\textwidth}
\centering
\caption{Core extraction performance (F1, Precision, Recall) across methods and tiers.}
\label{tab:core_performance_f1_precision_recall}
\begin{tabular}{l l c c c c}
\toprule
\textbf{Model} & \textbf{Method} & \textbf{Tier} & \textbf{F1} & \textbf{Precision} & \textbf{Recall} \\
\midrule
GPT-4.1-mini    & Prompt-based  & Simple   & 0.60 & 0.59 & 0.74 \\
                & RAG      & Simple   & \textbf{0.73} & 0.76 & 0.72 \\
\midrule
GPT-4.1-mini    & Prompt-based  & Medium   & 0.36 & 0.30 & 0.46 \\
                & RAG      & Medium   & \textbf{0.36} & 0.36 & 0.36 \\
\midrule
GPT-4.1-mini    & Prompt-based  & Complex  & \textbf{0.34} & 0.44 & 0.30 \\
                & RAG      & Complex  & 0.23 & 0.46 & 0.18 \\
\midrule
Mistral Small 3.2    & Prompt-based   & All tiers (avg.) & 0.34 & 0.35 & 0.41 \\
                & RAG      & All tiers (avg.) & \textbf{0.40} & 0.47 & 0.39 \\
\bottomrule
\end{tabular}
\end{subtable}

\vspace{0.75em}

\begin{subtable}[t]{\textwidth}
\centering
\caption{Inter-annotator agreement among human experts (10 pairwise comparisons).}
\label{tab:inter_annotator_summary}
\begin{tabular}{l c c}
\toprule
\textbf{Statistic} & \textbf{Jaccard} & \textbf{Cohen’s $\kappa$} \\
\midrule
Mean & 0.416 & 0.409 \\
Standard deviation & $\pm$0.189 & $\pm$0.258 \\
Range & [0.083, 0.714] & [-0.056, 0.746] \\
\bottomrule
\end{tabular}
\end{subtable}

\vspace{0.75em}

\begin{subtable}[t]{\textwidth}
\centering
\caption{Annotation performance: human experts vs. SecBPMN2 Assistant.}
\label{tab:human_vs_system_performance}
\begin{tabular}{l c c c}
\toprule
\textbf{Annotator} & \textbf{F1} & \textbf{Precision} & \textbf{Recall} \\
\midrule
Human expert average & 0.33 & 0.29 & 0.50 \\
SecBPMN2 Assistant (GPT-4.1-mini, RAG) 
                   & \textbf{0.52} & \textbf{0.58} & \textbf{0.52} \\
\bottomrule
\end{tabular}
\end{subtable}

\end{table}

\textit{\textbf{Structural Validity and Extraction Performance (E1--E2).}} Table~\ref{tab:schema_validity_by_complexity} reports schema validity of generated SecBPMN2 annotations after mapping and validation. Schema validity is evaluated first, as it indicates whether the pipeline produces well-formed and attachable security annotations. Across all complexity tiers, both GPT-4.1-mini and Mistral Small 3.2 achieve consistently high validity, confirming that the hybrid extraction–mapping pipeline reliably enforces SecBPMN2 structural constraints. RAG yields the strongest results overall, with GPT-4.1-mini attaining the highest validity on simple (0.987), medium (0.911), and complex (0.924) models, demonstrating that structural correctness is preserved as process complexity increases.

Table~\ref{tab:core_performance_f1_precision_recall} reports F1, precision, and recall across models, extraction strategies, and complexity tiers. Extraction performance declines with increasing model complexity, reflecting the richer and more heterogeneous security annotations of multi-actor processes. For GPT-4.1-mini, RAG achieves the highest F1 on simple workflows (0.73), while prompt-based extraction performs better on complex models (0.34 vs. 0.23). On medium-tier processes, both strategies perform comparably. For Mistral Small 3.2, results are reported as averages across tiers; although overall F1 remains lower than GPT-4.1-mini, RAG improves performance (0.40 vs. 0.34), indicating that lightweight models benefit from external context.

Interestingly, retrieval augmentation is not uniformly beneficial across process complexity tiers. While RAG substantially improves performance on simple models, GPT-4.1-mini achieves lower F1 on complex processes than the prompt-based configuration. A possible explanation is that, as process complexity increases, the retrieved SecBPMN2 and STS-ML documentation adds contextual information that competes with process-specific cues required for accurate annotation recovery. Prior work has shown that language models often struggle to effectively utilize relevant information distributed within long contexts and may underuse information embedded among large amounts of surrounding content \cite{liu2024lost}. Consequently, retrieval may encourage more conservative annotation decisions in complex models, reducing the number of recovered annotations despite improving schema grounding. More generally, the observed performance trend suggests that extraction difficulty is influenced not only by process size but also by the composition of the annotation task. In particular, medium and complex models contain a more diverse mix of annotation targets, including activities, data objects, and message flows, which increases the complexity of mapping security requirements to schema-compatible BPMN elements.

Precision–recall trade-offs reveal complementary failure modes. On complex processes, GPT-4.1-mini favors precision over recall, omitting a non-trivial fraction of intended annotations, which in security-by-design settings corresponds to unmitigated risks. Mistral Small 3.2 exhibits a more balanced precision–recall profile, with RAG favoring precision over recall and producing more conservative but structurally cleaner annotations. While this reduces over-specification, it may still result in missed annotations, increasing the need for expert review. These results indicate that GPT-based configurations are preferable when minimizing omissions is paramount, whereas Mistral Small 3.2 is better suited to scenarios prioritizing precision and controlled annotation generation.

\textit{\textbf{Human Baseline Comparison (E3).}} Table~\ref{tab:inter_annotator_summary} reports inter-annotator agreement and human–system performance. The human baseline was constructed from annotations independently produced by three experts with prior SecBPMN modeling experience. Each annotator was randomly assigned ten process models spanning simple, medium, and complex tiers to ensure balanced coverage across different process complexity. Human annotations exhibit moderate agreement (mean Jaccard 0.416, mean Cohen’s $\kappa$ 0.409), indicating that the task is challenging but yields a reasonable level of consistency among practitioners. The remaining variation reflects the interpretive nature of mapping security requirements to BPMN elements, particularly in complex models. Human annotators required on average nine minutes per process (ranging from approximately two minutes for simple workflows to over twenty minutes for complex models). Against this baseline, the SecBPMN2 Assistant achieves higher F1 and precision with comparable recall while requiring substantially less annotation time. The higher precision further suggests that the system generates fewer semantically incorrect or unsupported annotations than the average human annotator.

\begin{figure} [t]
\centering
\includegraphics[width=\textwidth]{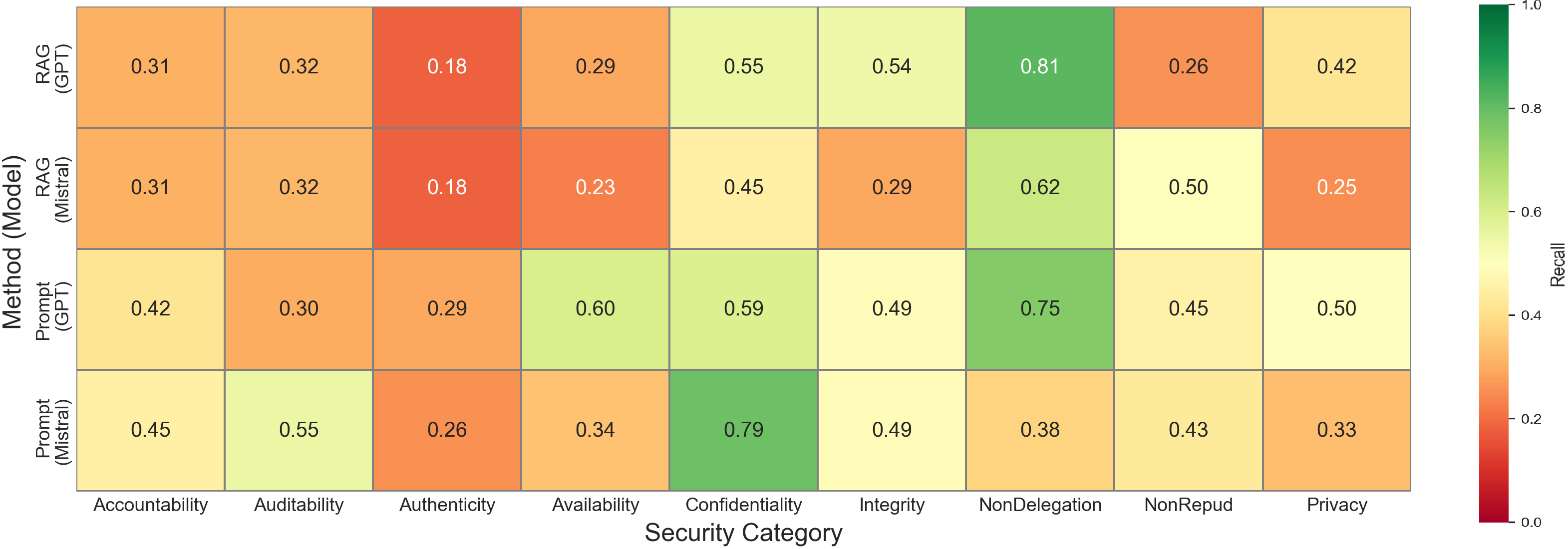}
\caption{GPT and Mistral category-level recall per security goal (Prompt-based, RAG).}
\label{fig:category_recall_heatmap}
\end{figure}

\textit{\textbf{Category Analysis (E4).}} Figure~\ref{fig:category_recall_heatmap} shows category-level recall across security goals. Recall is the highest for CIA-triad (confidentiality, integrity, and availability) objectives up to 0.79, which are frequent and typically expressed explicitly in text. By contrast, goals such as authenticity, accountability, and auditability occur less often and are commonly expressed implicitly, resulting in lower recall. This pattern indicates that the pipeline reliably captures protection-oriented requirements, while assurance- and compliance-oriented annotations benefit from expert oversight.

\textit{\textbf{Token Cost, Runtime, and Scalability (E5).}} GPT-4.1-mini consistently uses more tokens ($\approx$18.6k–20.6k per run) and incurs higher latency than Mistral Small 3.2 ($\approx$15.7k–19.4k tokens), reflecting its larger capacity and broader reasoning. RAG further increases token usage and latency due to retrieved context and longer outputs. For both models, latency scales predictably with process complexity and extraction strategy. In deployment, Mistral Small 3.2 enables faster, lightweight inference for interactive or high-throughput scenarios, whereas GPT-4.1-mini trades higher computational cost for greater accuracy and closer alignment with human annotations.

Overall, the results indicate that the proposed pipeline satisfies the evaluation objectives. It consistently produces structurally valid SecBPMN2 annotations (E1), achieves competitive extraction accuracy relative to expert annotations (E2–E3), captures the majority of explicit security requirements across goal categories (E4), and scales predictably with process complexity while maintaining practical computational cost (E5).

\section{Discussion and Conclusion}
\label{conclusion}

This paper introduced a hybrid LLM- and rule-based pipeline for automatically deriving SecBPMN2 security annotations from natural-language process descriptions. The approach combines semantic interpretation of textual requirements with rule-based enforcement of SecBPMN2 schema constraints, enabling the generation of structurally valid security-aware BPMN models. By decoupling semantic extraction from schema validation, the pipeline reconciles the flexibility of natural-language specifications with the strict structural requirements of security-oriented process modeling.

An empirical evaluation on 27 curated text–model pairs shows that schema-aware prompting and compatibility-constrained mapping improve annotation quality across models and extraction strategies. The pipeline consistently produces SecBPMN2-compliant annotations, achieving higher precision than human annotators while maintaining comparable recall with substantially less manual effort. The low inter-annotator agreement among human experts highlights the inherent ambiguity and cognitive burden of manual security modeling, underscoring the practical value of automated, schema-constrained assistance. Category-level analysis further indicates that the pipeline performs best for CIA-triad goals, such as confidentiality, integrity, and availability, whereas assurance- and compliance-oriented annotations, including accountability and auditability, remain more challenging and benefit from expert review.

Several limitations should be considered. Support for the full range of BPMN constructs remains partial, as the normalized representation abstracts away certain modeling details. While structurally inconsistent or non-attachable annotations can be detected and filtered, the system does not yet support automatic repair or conflict resolution. In addition, the evaluation is based on a curated dataset of 27 expert-annotated process models aggregated from the literature. Although the dataset spans multiple domains and was systematically aligned with textual security descriptions, its size reflects the current scarcity of publicly available SecBPMN2 benchmarks and may limit the generalizability of the findings.

Future work will focus on improving the accuracy and robustness of the proposed pipeline through larger benchmark datasets and the evaluation of the latest foundation models. We also plan to investigate retrieval strategies for complex process models, including methods for improving the relevance and utilization of retrieved security knowledge in long-context settings, as well as alternative intermediate representations such as graph-based encodings and their impact on token efficiency, annotation quality, and scalability. Furthermore, a more detailed analysis of the relationship between process complexity, runtime, token consumption, and computational cost would provide deeper insights into the scalability characteristics of the approach. Another promising direction concerns collaborative security modeling, where multiple stakeholders contribute requirements that must be consolidated into a shared process model, including mechanisms for detecting and resolving conflicts between security annotations.

\bibliographystyle{splncs04}
\bibliography{mybibliography}
\end{document}